\documentclass{elsart5p}

\usepackage{graphicx}
\usepackage{epsfig}

\usepackage{amssymb}

\begin{document}

\begin{frontmatter}

\title{Measuring the Neutron Lifetime Using Magnetically Trapped Neutrons}
\author[label1]{C.M. O'Shaughnessy,}
\author[label1]{R. Golub,}
\author[label1]{K.W. Schelhammer,}
\author[label1]{C.M. Swank,}
\author[label1]{P.-N Seo,}
\author[label1]{P.R. Huffman\corauthref{cor1},}
\ead{Paul\_Huffman@ncsu.edu}
\author[label2]{S.N. Dzhosyuk,}
\author[label2]{C.E.H. Mattoni,}
\author[label2]{L. Yang,}
\author[label2]{J.M. Doyle,}
\author[label3]{K.J. Coakley,}
\author[label3]{A.K. Thompson,}
\author[label3]{H.P. Mumm,}
\author[label4]{S.K. Lamoreaux,}
\author[label5]{G. Yang}
\address[label1]{North Carolina State University, 2401 Stinson Drive,
Raleigh, NC USA}
\address[label2]{Harvard University, 17 Oxford Street, Cambridge, MA USA}
\address[label3]{National Institute of Standards and Technology, 100
Bureau Drive, Gaithersburg, MD USA}
\address[label4]{Yale University, 217 Prospect Street, New Haven, CT USA}
\address[label5]{University of Maryland,College Park, MD USA}
\corauth[cor1]{North Carolina State University, Physics Department, Raleigh, NC
27695-8202 USA}

\begin{abstract}

The neutron beta-decay lifetime plays an important role both in understanding
weak interactions within the framework of the Standard Model and in theoretical
predictions of the primordial abundance of $^4$He in Big Bang Nucleosynthesis.
In previous work, we successfully demonstrated the trapping of ultracold
neutrons (UCN) in a conservative potential magnetic trap.  A major upgrade of
the apparatus is nearing completion at the National Institute of Standards and
Technology Center for Neutron Research (NCNR).  In our approach, a beam of
0.89~nm neutrons is incident on a superfluid $^4$He target within the minimum
field region of an Ioffe-type magnetic trap.  A fraction of the neutrons is
downscattered in the helium to energies $<200$~neV, and those in the
appropriate spin state become trapped.  The inverse process is suppressed by
the low phonon density of helium at temperatures less than 200~mK, allowing the
neutron to travel undisturbed.  When the neutron decays the energetic electron
ionizes the helium, producing scintillation light that is detected using
photomultiplier tubes.  Statistical limitations of the previous apparatus will
be alleviated by significant increases in field strength and trap volume
resulting in twenty times more trapped neutrons.

\end{abstract}

\end{frontmatter}

\section{Introduction}

The neutron lifetime is an important parameter in models of Big Bang
Nucleosynthesis (BBN) as well as other aspects of physics including the number
of neutrino flavors.  Along with precise measurements of the neutron beta-decay
correlation coefficients, the neutron lifetime can help to experimentally
determine the first element in the Cabibbo-Kobayashi-Maskawa (CKM) matrix,
$V_{ud}$.  Precision measurements of the first row yield a strong test
of the unitarity of this matrix.  See recent review articles \cite{Abe08},
\cite{Nic05a}, and \cite{Sev05} for a complete overview of the importance of
the neutron lifetime.

A currently published average value of the neutron lifetime, ($885.7 \pm
0.8$)~s~\cite{PDG}, is composed of seven statistically consistent measurements
using two techniques commonly referred to as ``beam'' and ``material bottle''
measurements.  A single material bottle measurement, however, dominates the
precision of this average.  Additionally, the most recent material bottle
measurement has comparable precision~\cite{Serebrov05} but is not included in
this average as it is more than six standard deviations discrepant from the
Particle Data Group's value.  This paper describes the development of a third
technique that combines magnetic confinement of neutrons with event-by-event
detection of decays~\cite{Huffman00NAT,Brome01}.  This new approach has a
fundamentally different set of systematic uncertainties than those of previous
work and offers the promise of comparable statistical precision as the recent
material bottle results.  As such, it should play a significant role in the
resolution of the current experimental situation.

\section{Experimental Method}

Our method utilizes ultracold neutrons (UCN) confined within a
three-dimensional magnetic trap.  Energy dissipation of the neutrons must occur
within the trapping region due to the conservative nature of the trap.  This
occurs when 12~K neutrons (0.89~nm) downscatter in superfluid $^{4}$He to near
rest via single phonon emission~\cite{Golub75}.  The UCN then interact with the
magnetic field via their magnetic moment, $\mu_n$.  For the neutron, $\mu_n$ is
anti-parallel to its spin, thus when the spin is aligned with the magnetic
field, it will seek to minimize its potential energy by moving towards low
field regions.  The energy of the downscattered neutrons is low enough that the
spin adiabatically follows the direction of the magnetic field throughout the
orbit.  Thus UCN with energies below the trap depth and in the
low-field-seeking state are trapped.

The UCN population is thermally detached from the helium bath allowing
accumulation of UCN to a density as high as $P\tau$, where $P$ is the
superthermal production rate and $\tau$ is the UCN lifetime in the source.
Neutron decay events are detected by turning off the cold neutron beam and
observing the scintillation light created by the beta-decay electrons.  When an
electron moves through liquid helium, it ionizes helium atoms along its track.
These helium ions quickly recombine into metastable He$_2^*$ molecules that are
in excited states.  About 35~\% of the initial electron energy goes into the
production of extreme ultraviolet (EUV) photons from singlet decays, yielding
approximately 22~photons/keV\@.  These EUV photons are frequency down-converted
to blue photons using the organic fluor tetraphenyl-butadiene (TPB) coated onto
a diffuse reflector surrounding the trapping region.  This light is transported
via optics to room temperature and detected by two photomultiplier tubes
operating in coincidence.  Our detection method allows us to observe neutron
decay events \emph{in~situ} and therefore directly measure the full decay
curve.

\begin{figure}[t]
\begin{center}
	\includegraphics[width=0.35\textwidth]{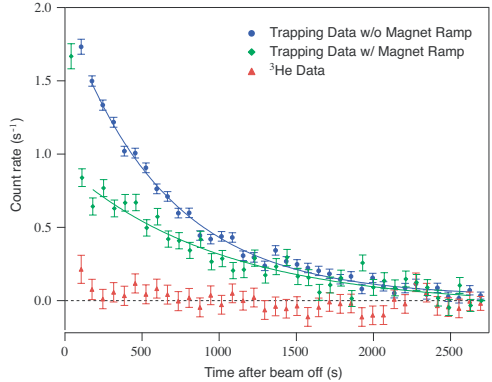}
\end{center}
\caption{Background subtracted neutron decay data taken at NIST.  The upper
curve represents data taken with  a constant magnetic field.  The middle
curve is data taken after the trap field is lowered briefly to remove the above
threshold neutrons.  The lower curve is data taken with natural isotopic
abundance helium, where one would expect no trapped neutrons due to capture on
$^3$He.} 
\label{fig:data}
\end{figure}

Using our prototype apparatus, neutron trapping data was taken in a number of
different experimental configurations primarily for investigating systematic
effects.  This data is discussed in detail in Ref.~\cite{Dzho04} and shown in
Fig.~\ref{fig:data}.  The raw neutron decay curve exhibits a trap lifetime of
$621^{+18}_{-17}~(stat)$~s.  The low central value suggests that there
exist neutron trap loss mechanisms other than beta decay.  These losses are a
result of the presence of above threshold neutrons.  We have shown that these
neutrons can be removed from the trap by briefly lowering the magnetic field
immediately after the loading period, causing the neutron orbits to expand and
intersect the material walls where absorption can occur.  As seen in
Fig.~\ref{fig:data}, data collected using this field-ramping procedure exhibits
a lifetime consistent with the world average, $\tau =
831^{+58}_{-51}~(stat)$~s, albeit with a large statistical uncertainty.
In this data, the field was lowered to 30~\% of its maximum value immediately
after the neutron beam was turned off.  In addition to any above threshold
neutrons, the ramp flushes approximately 50~\% of the fully trapped neutrons
from the trap.

Using these results and our continuing studies of systematic effects, a next
generation apparatus has been designed and is close to being operational.  Our
experimental exploration of systematic effects using the previous apparatus was
severely limited by counting statistics. However, of the significant systematic
uncertainties, including marginally trapped neutrons, $^3$He contamination,
gain drift, and background subtraction, the largest were of the same order as
the statistical precision. Additional studies with that apparatus would not
have been productive.  For example, each systematic study required
approximately 7-10~days of beam-time and the entire set of data took close to
one year to collect.  It was essential that we incorporate our previously
developed KEK trap into the apparatus to move forward.

\section{Newly Developed Apparatus}

\begin{figure}[h]
\begin{center}
	\includegraphics[width=3in]{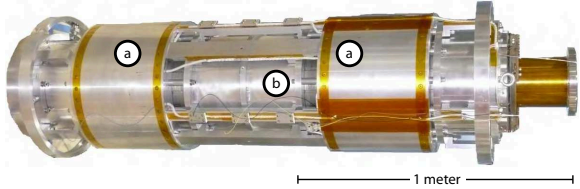} \hspace{0.5in}
\end{center}
\caption{Photograph of the KEK magnetic trap showing (a) the solenoids, (b) the
quadrupole within an aluminum support.}
\label{fig:magnet}
\end{figure}

At the heart of our new apparatus is a high-current Ioffe-type magnetic trap.
This trap consists of an accelerator-type superconducting quadrupole magnet
that provides radial confinement combined with two aligned low-current
solenoids that provide axial confinement.  The quadrupole magnet is on loan
from the High Energy Accelerator Research Organization (KEK) laboratory in
Japan while the two solenoids were designed and constructed in collaboration
with American Magnetics Inc.  The new trap has a depth of 3.1~T and trapping
volume of 8~l.  A complete description of this and our lower-current prototype
traps can be found in Ref.~\cite{Yang08} and a photograph of the KEK trap is
shown in Fig.~\ref{fig:magnet}.
The quadrupole operates at a current of 3400~A and the solenoids at 250~A, with
both magnets protected with either silicon-controlled rectifiers (SCR) or
diodes.  The protection for the quadrupole, for example, uses SCRs coupled to a
dump resistor to attenuate 95~\% of the stored energy within 1.5~s.  The
current from the room-temperature power supplies is transported into the
cryogenic helium bath using a pair of 5000~A high-temperature superconducting
(HTS) current leads \cite{Lock69}.  These leads significantly reduce the heat
loads to the cryostat as compared to conventional vapor-cooled leads.  A second
set of HTS leads are used for the 250~A entry.  To additionally reduce helium
consumption, we incorporated into the apparatus two cryocoolers, each with
1.5~W of cooling power at 4.2~K.  The total liquid helium consumption of the
apparatus is held to below 100~l/day.

\begin{figure}[h]
\begin{center}
	\includegraphics[width=2.5in]{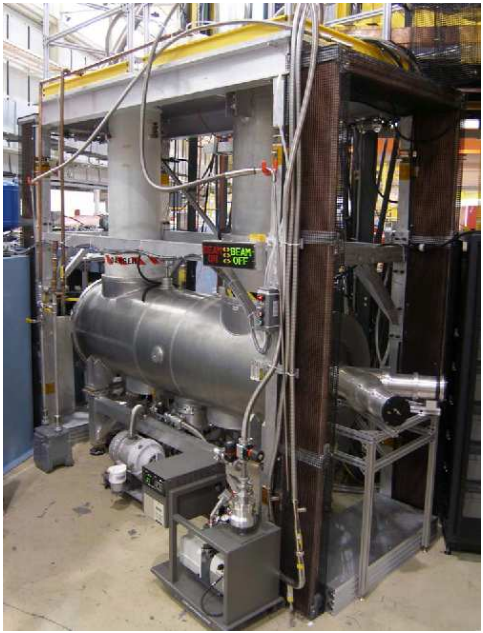} \hspace{0.5in}
\end{center}
\caption{The new apparatus on the NCNR beamline. The neutron beam enters from
the left; the light detection system can be seen at the center-right.}
\label{fig:dewar}
\end{figure}

The trap assembly has been tested in the new cryostat (Fig.~\ref{fig:dewar}) and
reached $> 85$~\% of its design value before the first quench.  Successive
magnet training was not possible on this first attempt due to a malfunction in
one of the cryocoolers.  However, in a separate test with the magnet oriented vertically,
the magnet was trained to the limit of the vapor-cooled current leads in the
dewar, reaching $>~90$~\% of the load line.

\begin{figure}[h]
\begin{center}
	\includegraphics[width=3in]{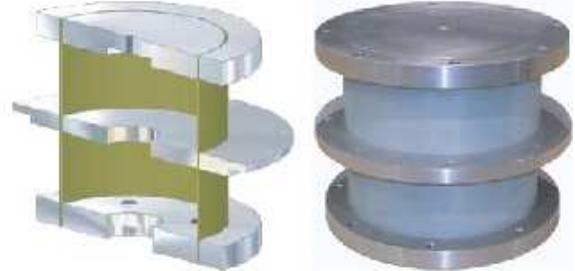} \hspace{0.5in}
\end{center}
\caption{G-10 fiberglass/aluminum support post for the Ioffe magnetic trap;
cross-sectional schematic (left) and a photograph of one of the posts (right).}
\label{fig:posts}
\end{figure}

To mechanically support the magnetic trap while minimizing the heat-load into
the liquid helium bath, we designed and tested two cryogenic support
posts~(Fig.~\ref{fig:posts}) based on a design by T. Nicol \emph{et~al.} at
Fermilab~\cite{Nicol:1987mq}.  The body of each post is a 19~cm long, 19~cm
diameter G-10 fiberglass tube with 1~mm wall thickness.  Equally spaced
aluminum flanges at 300~K, 77~K, and 4.2~K provide lateral mechanical support
for the tube and bolt holes for attachments to the dewar.  The posts were load
tested with an active load of 1360~kg at room temperature.  The heat-load from
each post is calculated to be $\sim$ 0.35~W at 4.2~K\@.

A ``U''-shaped dewar houses the magnetic trap and is shown in
Fig.~\ref{fig:dewar}.  On the beam-entrance end, the vertical tower houses an
Oxford\footnote{Any mention of commercial products or reference to commercial
organizations is for information only; it does not imply recommendation or
endorsement by NIST nor does it imply that the products mentioned are
necessarily the best available for the purpose.} 400 $^{3}$He-$^{4}$He dilution
refrigerator that is used for cooling the helium-filled target cell.  The
second tower houses the current leads for the magnets and provides a liquid
helium reservoir for the superconducting magnets.

Neutrons enter the trapping region through a series of perfluoroalkoxy (PFA)
Teflon/beryllium windows on the dewar~\cite{Butterworth98}.  Interlocking tubes
of boron nitride (BN) surround the beam and shield the dewar from scattered
neutrons to minimize backgrounds from neutron-induced activation.  The
remaining beam is absorbed at the end of the cell by the acrylic light guide.
We have shown that the neutrons absorbed in the acrylic do not cause
significant color center formation at our present fluences and hence will not
degrade the light detection efficiency~\cite{Mattoni02}.  Non-trapped neutrons
scattered within the helium-filled trapping region are absorbed by high-purity
BN shielding (3mm wall) exterior to the light detection system.  Thin carbon
tubes (1mm wall) line the BN to absorb any neutron-induced luminescence
originating from color center formation in the BN.  The helium is contained in
a stainless steel (SS) tube that extends through the length of the magnet and
houses the shielding and light guides.  Thermal isolation of the 25 kg cell
assembly is achieved by suspension at each end by three pre-tensioned 2.6~mg/cm
Zylon yarns.  

The trapping region contains isotopically pure $^{4}$He (less than
$5\times10^{-16}$ parts $^{3}$He~\cite{McClintock78})\footnote{For more
information on Accelerator Mass Spectrometry~(AMS) measurements to directly
verify the isotopic purity see Ref.~\cite{Yang06}.}that is cooled to $T <
200$~mK to reduce phonon upscattering.  The helium in the trapping region is
thermally connected to the dilution refrigerator using a superfluid helium heat
link as opposed to copper links.  In addition to acting as a helium fill line,
the helium heat link minimizes eddy-current heating in the cell when the
magnetic field strength is changed.  

The light detection system consists of a 11.4~cm outer diameter
Goretex tube with a thin layer of TPB evaporated onto the inner
surface.  The end of the tube is viewed by a 11.4~cm diameter acrylic rod that
transports the light outside the magnet bore.  The light exits the $<200$~mK
region, passes through a 0.6~cm thick acrylic window at 4.2~K, and then into a
14~cm diameter acrylic rod that transports the light outside of the apparatus.
Once exiting the vacuum region, the diameter of this light guide is reduced to
11.4~cm in order to better couple with downstream photomultiplier tubes (PMTs)
through a Winston cone optimized for this geometry~\cite{Welford89}.  The light
is then divided using a Y-shaped acrylic splitter to couple the light into two
12.7~cm diameter PMTs that are operated in coincidence mode to reduce
uncorrelated low-temperature luminescence and dark-current backgrounds.

Pulses are digitized using transient waveform digitizers and stored for
off-line analysis.  This allows one to vary lower-level thresholds in software.
Background events arising from cosmic rays passing through the acrylic and
helium are actively vetoed using an array of scintillation paddles that
surround the lower half of the apparatus.  Background radiation from
instruments in the surrounding area could also introduce background events, so
the whole lower section of the apparatus is shielded by 10~cm of lead and 5~cm
of polyethylene at room temperature.  The remaining backgrounds arise from
neutron-induced activation of materials within the detection region,
neutron-induced luminescence, and gamma-rays Compton scattering within the
acrylic.

\section{Improvements and Future Directions}

Data taken using our previous apparatus (Fig.~\ref{fig:data}) indicated
that we were trapping approximately 3000 neutrons per fill cycle.  Once the
magnetic field ramping procedure was complete, approximately 1600 neutrons
remained in the trap.  Scaling to both the larger volume and greater depth of
the new KEK trap, the number of trapped neutrons at NIST is expected be
approximately $4.5 \times 10^{4}$, or about 5~cm$^{-3}$, after a similar 30~\%
field ramp.  Assuming that the overall detection efficiency shows the same
behavior as a function of position within the trap, the observed decay rate
will be increased by a factor of $\sim 20$, allowing systematic tests to be
performed in days now as opposed to the weeks required previously.

The UCN production rates have been calculated using a Monte Carlo simulation
that begins with the polychromatic neutron beam, incorporates Bragg diffraction
from the 0.89~nm neutron monochromator, and propagates the 0.89~nm neutrons
into the measurement cell.  With this code, we were able to reliably predict
the number of trapped neutrons in the previous apparatus~\cite{Mattoni02}.
Estimates made at the early stages of this project did not fully incorporate
the loss of sensitivity due to the higher-than-expected background counting
rates and were thus overly optimistic.

The overall design and composition of the new apparatus has not changed
significantly.  That is, the cell construction, detection system, cooling
system, and neutron optics are scaled versions of the previous design.  This
gives us high confidence in the predictions of the number of neutrons trapped
as well as the detection efficiency for the new apparatus.  For the sensitivity
estimates, we have assumed both that the backgrounds scale appropriately with the
volumes of the various components of the system and that detection
efficiencies have remained the same even though improvements in both of these
systems are being implemented.

\begin{figure}[h]
\begin{center}
	\includegraphics[width=3in]{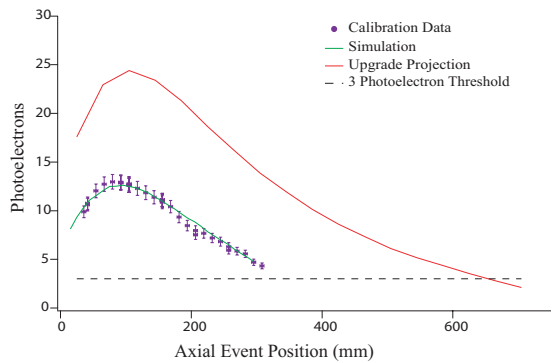} \hspace{0.5in}
\end{center}
\caption{Average number of photoelectrons per decay.  Measured (points) and
calculated (lines) as a function of position along the cell axis.}
\label{fig:efficiency}
\end{figure}

An experimental calibration of the detection efficiency using a 364~keV
$^{113}$Sn beta line source was performed during the previous data collection
run and has been used as a benchmark for optimizing our light collection system
using the CERN Monte Carlo code called GuideIt.  Modifications to this code
were implemented to more realistically generate a photon source distribution in
the modeled TPB geometry due to a point source of EUV photons.  In reality this
will be along ionization tracks in the helium with lengths on the order of a
centimeter.  However a point source satisfactorily reproduces the data
(Fig.~\ref{fig:efficiency}).

It was determined that two components of the previous light collection system
resulted in excessive attenuation of photons and have been eliminated in the
current design; the B$_{2}$O$_{3}$ beam stop and a sapphire window at 77~K\@.
The beam stop was originally installed to minimize any activation and color
center formation in the acrylic light guides.  In independent tests, we have
determined that neither of these issues are a problem and thus have removed
this component in the new design.  The sapphire window was included to provide
good thermal conductivity for cooling the 77~K light guide, thereby reducing
the heat-loads to the experimental cell.  In the new design, we have extended
the distance between the 77~K and 300~K end flanges such that a longer guide
can be used.  Based on results from an ANSYS finite element analysis of the
heat flow in the new 77~K guide, the sapphire will no longer be needed.  From
these improvements and optimizations, we expect to gain a factor of 1.9 in
average light collection efficiency, as can be seen in
Fig.~\ref{fig:efficiency}.  Such an improvement will allow us to increase the
lower discrimination levels and thus significantly reduce the component of the
signal arising from background events.

Additionally, new designs for thermal and mechanical clamps to the 77~K
light guide element will also improve performance of the new apparatus.  In
order to ensure the full thermal contact at 77~K and a vacuum seal at 300~K two
flanges were precisely machined to have an inner diameter smaller than the
outer diameter of the acrylic rod.  The rod was then cooled to 77~K such that
thermal contraction allowed it to slide through the room temperature flanges.
However, upon simultaneous cooling of the aluminum and the acrylic, the
dimensions are such that full contact of the mating surfaces will remain until
temperatures are well below 77~K\@.  Upon repeated vacuum, and cooldown cycling
this guide element has reliably performed its functions. 

The apparatus is presently on the beamline at the NCNR~(Fig.~\ref{fig:dewar}).
It has been successfully cooled, the magnet has been tested, helium has been
liquefied into the cell and cooled with the dilution refrigerator.  During our
next cooldown, the goal is to introduce neutrons into the trapping volume and
confirm the expected number of trapped neutrons.

In addition, we will continue to address our understanding of the largest
potential source of systematic uncertainty, marginally trapped or above
threshold neutrons.  A 3-dimensional numerical grid of the solenoidal and
quadrupole magnetic field was computed using the Radia plug-in to Mathematica
software.  Using a symplectic integration code~\cite{Yerozolimsky97}, neutrons
are then tracked through the resultant potentials in both cases where fields
are static, and where the quadrupole field is ramped down and the solenoid
field is fixed.  Based on an optical approximation for the material Fermi
potential~\cite{Golub91} at the TPB/Goretex, BN and graphite walls of the trap,
this model predicts the likelihood of the tracked neutron to be absorbed in a
material surface or to be reflected back into the cell region. Thus it is used
to determine the cumulative survival probability of each neutron as a function
of time. We plan to quantify the ratio of the integrated survival probabilities
of UCN above threshold and below threshold for various field ramping
strategies.  In an idealized theoretical study, we will quantify the systematic
error in the mean lifetime estimate as a function of the relative fraction of
surviving above threshold UCN and a decay time parameter for their survival in
the trap. As a physical grounding, neutron reflectometry measurements of the
cell wall materials will be performed, and these measured values will be
updated in the model.  

A significant benefit of the new apparatus is that the increased count rates
expected will give us ample time to perform a series of systematic checks such
as measuring the neutron loss due to phonon upscattering and a careful
characterization of the background subtraction. Additionally, a benchmark for
the model of above threshold neutrons could be obtained from the scaling of
measured trap lifetime as a function of field ramp parameters. Once these
systematic studies are successful, we will then pursue measuring the neutron
lifetime.

Assuming the same detection efficiency (although we expect a significant
improvement) and the same backgrounds from other instruments (scaled
appropriately with the increased volumes of the new apparatus), we anticipate
that we will be able to perform a measurement with a statistical precision
corresponding to $<2.4$~s in a 40~day reactor cycle.  With the improvements
discussed above, a more optimistic estimate places this value at 1.5~s.  Such a
measurement would play in important role in clarifying the current uncertainty
surrounding the neutron lifetime.

This work was supported by the National Science Foundation under Grant Nos.\
PHY-0354264 and PHY-0354263 and by NIST. The neutron facilities used in this
work were provided by the U.S. Department of Commerce.



\end{document}